# Axon Hillock Currents Allow Single-Neuron-Resolution 3-Dimensional Functional Neural Imaging Using Diamond Quantum Defect-Based Vector Magnetometry

Madhur Parashar[1], Kasturi Saha[2]*, Sharba Bandyopadhyay[3]*

[1]School of Medical Science and Technology, Indian Institute of Technology Kharagpur, Kharagpur 721302, India
[2]Department of Electrical Engineering, Indian Institute of Technology Bombay, Powai, Mumbai 400076, India
[3]Department of Electronics and Electrical Communication Engineering and Advanced Technology Development Centre, Indian Institute of Technology Kharagpur, Kharagpur 721302, India

*corresponding authors
Correspondence email: [3]* sharba.ban@gmail.com, [2]*kasturis@ee.iitb.ac.in

## Abstract

Magnetic field sensing, with its recent advances, is emerging as a viable alternative to measure functional activity of single neurons in the brain by sensing action potential associated magnetic fields (APMFs). Measurement of APMFs of large axons of worms have been possible due to their size. In the mammalian brain, axon sizes, their numbers and routes, restricts using such functional imaging methods. With segmented model of mammalian pyramidal neurons, we show that the APMF of intra-axonal currents in the axon hillock are two orders of magnitude larger than other neuronal locations. Expected 2-dimensional vector magnetic field maps of naturalistic spiking activity of a volume of neurons via widefield diamond-nitrogen-vacancy-center-magnetometry (DNVM) were simulated. A dictionary based matching pursuit type algorithm applied to the data using the axon-hillock's APMF signature allowed spatiotemporal reconstruction of APs in the volume of brain tissue at single cell resolution. Enhancement of APMF signals coupled with NVMM advances thus can potentially replace current functional brain mapping techniques.

**Introduction**

The development of in vivo two photon calcium imaging[1] and subsequent development of fast voltage/calcium sensors[2,3] of neuronal membrane potential, has allowed probing of local neuronal circuits[4–8] in the mammalian brain at single-cell and millisecond spatiotemporal resolution. The advent of this technique actively triggered the study of excitatory-inhibitory populations of neurons and their functional connectivity[9–11] in passive sensory and active behavioural state of an organism. Multiphoton functional neuronal imaging is limited to depths of maximum of 1 mm from the brain surface due to limits of optical penetration of deep tissue and scattering[1,5,12,13]. Therefore, local neuronal populations from deep areas of the brain, like the hippocampus, many regions of frontal cortex, amygdala can't be probed at local circuitry level, unless largely invasive and potentially damaging optical fibres are used.

Methods of ultrasensitive microscale magnetic field sensing[14–20] or single cell resolution functional magnetic resonance imaging[21,22] are being progressively developed to address this challenge. Diamond Nitrogen Vacancy Centers (NVC) have emerged as a class of ultrasensitive nanoscale magnetic field detectors that function at ambient temperature[23–25]. Additionally, the diamond NVC's inert chemical nature allows it to be placed very close to the biological tissue[26] allowing sensitive probing of biological magnetic fields. In this context, Barry[27] et al experimentally demonstrated measurement of worm axon action potential associated magnetic field (APMF), which was found to be ~600pT peak-to-peak in magnitude, using an ensemble of 2-Dimensional NVC layer in diamond. Notably, high sensitivity microscale magnetic field mapping has been made possible by preparation of high density NVC diamond samples with high intrinsic coherence[14]. Further, the directional orientation of different NVC along the tetrahedral axes is used to obtain vector direction of magnetic field[27,28]. APMF signal bandwidth falls in DC to few kilohertz bandwidth. DC field sensitivity in Diamond NVC experiments is rapidly improving towards quantum projection shot noise limit[27], at ~100 fT/$\sqrt{\text{Hz}}$, with the best DC sensitivity record to be measurement at 50 pT/$\sqrt{\text{Hz}}$ for ensemble vector magnetometry measurements[29]. With constantly improving DC field sensitivity, the community is expected to capture mammalian spike signals using diamond NVCs.

However, a challenge in probing a network of mammalian neurons in vitro or in vivo will be to reconstruct action potential timing and location of single neurons from diamond NVC magnetometry. Previous work in reconstruction of simulated APMF recordings via diamond NVC magnetometry has been restricted to simple models of passive conducting axons via filtering of noise in spatial frequency domain[27] and a Wiener filter-based reconstruction of axonal firing[30]. Also, theoretical work has been carried out to develop inverse filters[31–33] for reconstruction of two cylindrical axonal currents, where analytical expression for current density was known. To the best of our knowledge, no method has been developed that takes into account the complex geometry and physiology of cortical neurons, where analytical expression for intra-axonal currents can't be derived. Further, single action potential event detection from a time series of measurements must also be combined with spatial reconstruction, a necessary feature which is absent in previous studies.

In this work we address the reconstruction of spike location and timing for realistic mammalian cortical pyramidal neurons, comprising of soma, axon hillock region, axon initial segment and other regions, specifically with respect to the case of measuring 2-Dimensional vector magnetic field map via widefield diamond NVC magnetometry. We simulated voltage propagation in a realistic cortical pyramidal neuron model[34] and obtained intra-axonal current profiles during an action potential. These spatiotemporal current profiles were used to estimate vector magnetic field during an action potential. We found a 36pT peak-to-peak mammalian APMF magnitude, which is close to current limits of diamond NVC based DC magnetometers. Notably, we found that axon hillock contributes almost two orders more, as compared to other axonal regions, to the measured APMF estimate. This naturally occurring advantage simplifies the inverse problem from being equivalent to solving randomly oriented current carrying wires, where the location of ultra-small current keeps changing in 3D space as the action potential propagates over hundreds of microns, to primarily a set of ~ 10 μms sized axon hillock region, fixed in space and exhibiting localized current flow only when an AP occurs in the corresponding cell soma.

We propose an adaptation of dictionary based matching pursuit algorithm[35–38], to be applied on measurements from widefield diamond NVC magnetometry, for solving individual spike timings and locations in a 3D volume of randomly oriented pyramidal neurons. We show that near 70 percent correct reconstruction can be achieved for randomly oriented neurons arranged in 3D volume and also, for neurons arranged in a 2D plane parallel to diamond NVC layer. We find that our matching pursuit-based algorithm allows high noise resilience to the reconstruction. Further, we analyse the closest distance between a pair of cells that can be resolved by the algorithm. We find that nearby single neurons spiking near simultaneously with timing difference 1ms or more can be reliably resolved. Based on reconstruction errors, we infer that strongly correlated columns of the dictionary, due to similarity of magnetic field patterns formed by two closely located neurons, are main constraints to achieving perfect reconstruction.

**Results**

*Action potential associated magnetic field have highest contribution from intra-axonal axon hillock currents*

APMF of worm (marine fanworm *Myxicola infundibulum* and the North Atlantic longfin inshore squid *Loligo pealeii*) single axons were estimated to be ~600 pT in magnitude measured with ensemble diamond NVC imaging setup[27]. However, mammalian neurons have significantly smaller cross sectional diameter (~1 µm at nodes) and carry orders of magnitude less axonal currents than the worm axon. APMF estimates of mammalian neurons reported from computational studies are inconsistent and vary in the range ~1pT- 1nT[27,39,40]. Further, the contribution of currents in the different types of axonal segments, like axon hillock, nodes of Ranvier and others, to the final APMF has not been investigated. In an intact mammalian brain, detecting activity based on APMFs from axonal currents requires localization of the source at single neuron resolution. Only 2D measurements of magnetic field via widefield ensemble DNVCM would not be usable, as the 3D source reconstruction would be non-unique. We first address the question – are there specific signatures in the APMF of a neuron that can allow spatiotemporal reconstruction of the source? For this purpose, we consider a realistic cortical pyramidal neuron APMF and investigate the differential contribution of neuronal regions to the APMF magnitude.

The voltage propagation through the structures in a realistic cortical pyramidal neuron model[34] was simulated to study the contribution of the distinct types of sodium channels in initiation of the action potential. The model incorporated realistic geometry, physiological parameters and experimentally determined ion channel densities. The pyramidal neuron comprised of different segments namely cell soma, dendrites, axon hillock, action initial segment, unmyelinated axon, myelinated axon and nodes of Ranvier. The neuron was divided into iso-potential compartments and membrane potential dynamics across these compartments was governed by the cable theory[41–43] equation (Eqn. 1, Methods), and solved using the NEURON solver (see Methods). A step current injection was added at 'central' soma segments to make the neuron fire APs and the membrane potential for one AP was recorded. Voltage propagation across each segment in time for a single AP was imported from NEURON[40,44] and all further analyses were done in MATLAB[45]. AP initiation was observed in the most distal segment of the axon initial segment (Fig. 1A, AIS region). This result is consistent with previous studies, which show that AP originates in the axon initial segment, due to the presence of high density of sodium channels and high resistance of segments[34]. After AP initiation, a bi-directional propagation of AP, one in the direction of cell soma and the other in the direction of axon, is observed (Fig. 1A-C). The bi-directionality is in agreement with observations from experiments[34]. In order to estimate the mammalian neuron's APMF, intra-axonal currents across segments at each time instant were calculated. Only the intra-axonal currents were considered in calculating the APMF based on previous theoretical work[32,46,47] showing that the net magnetic field due to spiking in neurons is primarily determined by the intra-axonal currents. This assumption has also been experimentally demonstrated in SQUID based magnetic field measurements of frog sciatic nerve[48,49]. The intra-axonal current profiles in each segment (Fig. 1D) were calculated (Eqn. 2, Methods) based on the discrete version of the basic cable equation. We observe high intra-axonal current flow in the axon hillock region (Fig. 1D). Further, we analysed intra-axonal current flow in all regions of the neuron, as a fraction of current flow in the second node of Ranvier (Fig. 1E). The discontinuities observed in the traces are due to artificial jumps in the current flow caused by the compartmentalization and would be absent in the actual case. We found two orders of magnitude higher current flow in the most distal segment of the axon hillock as compared to the second node of Ranvier. This result implies that pixels on the diamond NV sensor that are on the perpendicular axis of the axon hillock will sense significantly higher magnitudes of APMF signatures. Comparisons of APMF magnitude and waveforms (calculated by Eqn. 3, Methods) between a point located vertically below the axon hillock to a point located vertically below the soma or axon terminal clearly implicate the axon hillock as the dominant contributor to the APMF (Fig. 1F). We quantify the mammalian APMF magnitude as the peak-to-peak magnetic field (Y component) measured at a point vertically below the axon hillock, which is 36pT at a distance of $20.50 \ \mu m$ from the longitudinal axis passing through the centre of the axon hillock. However, no experimental verification of the mammalian neuron APMF magnitude have been made yet to the best of our knowledge.

*2-D DNVMM comprises of specific signatures of APMF*

We generated 2-D time varying magnetic field maps by spatially summing magnetic field contributions from current flow in different segments of the neuron, at each time instance (Eqn. 3, Methods). We explain the features in these maps with respect to the following: bidirectional propagation of action potential (AP), the activity in nodes of Ranvier and the overall spatial size of APMF signatures. These 2-D DNVMMs, are simulated realizations of magnetic field measurements on a 2D plane, as it would be in an experimental case of a thin top layer of NV defects in a cube of diamond (see Methods). Henceforth we refer to the above 2D DNVMMs as maps, unless mentioned otherwise. The simulated maps at different time points (Fig. 1A, arrows, i-vi) acquired during firing of an AP show a number of variations of features (Fig. 2). Here, we show noiseless 2D maps to understand fundamental features and correlate it with AP propagation in neurons. The first prominent feature in the maps is the dominance of the axon hillock, as we observed that APMF signatures are significantly visible in only frames iii-v (Fig. 2A), which correspond to AP propagating through or near the axon hillock region of the neuron (Fig. 2B). Since, the axon hillock activity's contribution to the APMF was approximately two orders larger than those of other regions APMF, signatures of other regions were unidentifiable in these maps (Fig. 2A, first panel of frames). Therefore, to understand the APMF signatures of other regions in these maps, we saturate the color axis at lower magnetic fields and separately analyze the $B_y$ and $B_z$ (second and third panels respectively of every frame) component of the magnetic field. Each map (Fig. 2A, i-vi) corresponds to AP propagating through specific segments (panels in Fig. 2B) and at specific time points (dashed vertical lines, Fig. 2B). The membrane potential of the three example segments (Fig. 2B, Axon Hillock, AIS and node of Ranvier) show that the AP initiation falls between frames ii and iii. After, initiation of the AP we observed bi-directional propagation of AP along the cell soma and axon terminals (Figs. 1A&2A). Later in time (4-6ms, after the start of current injection), we observed repetitive patterns of activity which correspond to AP propagation through repetitions of myelin-node in the neuron (iv-vi, Fig. 2A). Another important feature was the appearance of quadrant like $B_z$ components of the field, which indicates that intra-axonal currents can be approximated as current dipoles.

### *Formulating the DNVMM inverse problem as a dictionary based linear inverse problem*

The inverse problem comprises of detecting time and location of neurons that fired an AP from NV maps. The magnetic field, as measured by ensemble diamond magnetometry, due to current flow across segments of spiking neurons is given by Eqn. i.

$$B_{nv}(\alpha, \beta, t) = \sum_n k\, I(x, y, z, t)\, \frac{\overline{dl} \times \overline{dr}}{|\overline{r}|^2} \qquad (i)$$

In the above equation, $B_{nv}(\alpha, \beta, t)$ is the field experience by an NV centre or a small ensemble of NV centres located in pixel at position $(\alpha, \beta)$ at time t, n denotes all isopotential segments (of all neurons), k is a constant (Methods). The inverse problem is to detect a fraction of $I(x, y, z, t)$ waveform that guarantees an action potential in neuronal soma at location x, y, z and time t, by operating on the vector magnetic field $B_{nv}(\alpha, \beta, t)$ from different diamond NVC pixels obtained from diamond NVC vector magnetometry. It is to be noted that, since an AP is an all or none event, we don't need to reconstruct the full spatiotemporal variation of current I.

The above inverse problem can be formulated in terms of a dictionary based linear inverse problem, where the dictionary elements contain prior information about DNVMMs from AP firing of single neurons located at different spatial locations in different orientations. Since the Axon Hillock currents provide the dominant signature in a neuron's APMF, the individual dictionary elements are created by considering mainly the Axon Hillock DNVMMs (frames iii-v, Fig. 2A). The above linear representation allows application of a matching pursuit algorithm for spatiotemporal AP reconstruction[37,50,35,36]. A dictionary based matching pursuit approach is motivated by sparse spatiotemporal distribution of spikes in mammalian cortices[50-52] and previous application of matching pursuit algorithms to MEG/EEG data[53] for reconstruction of active current dipoles formed during APs. However, reconstruction of MEG/EEG have been demonstrated only at coarse spatial resolution, in the range of hundreds of microns, not near the single cell spiking resolution.

The final experimental map is expressed as a linear combination of individual DNVMMs. For the dictionary matrix *A*, *X* a vector of length equal to number of neurons and *B* a vector of experimental multidimensional DNVMM data points, we can write the problem as in Eqn. ii in which *X*, a binary vector, needs to be estimated.

$$AX = B + \epsilon \qquad (ii)$$

The dictionary *A* is of *m* x *n* dimensions, where *m* is the number of dimensions in the experimental data, $\epsilon$ is the noise in DNVMM experimental maps and *n* is the total number of neurons in the tissue volume of interest. We aim to solve for *X*, whose elements can only be zero or one depending on whether the corresponding neuron fired or not. However, the above linear equation is over determined, with the number of neurons being less than the number of dimensions in an NVMM. Performing a full least square search for $2^n$, where *n* is in thousand range, is computationally impractical and hence the following dictionary based matching pursuit algorithm is used after modification for DNVMM time series data.

### *Proposed matching pursuit algorithm*

The proposed reconstruction algorithm (details in Methods) works by considering a multidimensional time series ($B_t$) considering each pixel in maps of $B_x$, $B_y$ and $B_z$ at each of 3 successive time points (*t*-2, *t*-1, *t* for $n_{tp} = 3$ in Methods, Fig. 3A) as a dimension. Each of the $B_x$, $B_y$ and $B_z$ 2-D maps (image data representing each time point) in the time series are represented in Fig. 3A separated by white dashed lines. At each time instant, the signal acquired along with the signal in the previous 2 time frames can be projected onto individual normalized columns ($\hat{A}_t$) of the dictionary *A* (Fig. 3B). The maximum projection, and hence, the most probable single neuron that fired an AP at that instant is denoted by best matched neuron index in schematic Fig. 3. We impose the condition of detecting the spikes of a particular neuron as occurrences of the same neuron as the best matched neuron at multiple time instances, greater than a specified parameter $p1$ within a stretch of successive $p2$ time points. On detecting a spike, we ascertain the exact spike timing of the neuron by matching the neuron's spike signal to different regions of the experimental signal with a local shift of $+/- \tau$ timepoints (1.5 milliseconds total) near the time point where we detected neuronal spike. The time instant that gave maximum dot product/alignment with the neuron's signal was chosen at the exact spike timing of the particular neuron. Since axon hillock activity is the dominant signature on NVMMs, each element in the dictionary is constructed from additions of three time points 3.5ms, 4.0ms and 4.5ms, which mainly correspond to high axon hillock activity (Fig. 3B). The main control parameters of the reconstruction algorithm are threshold $T$, $p1$ and $p2$. The threshold needs to be set greater than the smallest energy column of the dictionary. $p2$ parameter depends on total number of time points considered in formation of the dictionary. $p1$ controls minimum number of consecutive time points that a neuron should be best matched to the signal to be considered a spiking event. For later demonstrations of reconstructions, we considered $p2$ equal to 3, which is equivalent to considering 3.5ms, 4ms and 4.5ms time point NVMMs being incorporated into the dictionary. $p1$ has been taken to be 2, equivalent to a millisecond length signal in real time.

### *Population performance in 2-Dimensional and 3-Dimensional cortical tissue simulations*

The feasibility of spatiotemporal localization of occurrence of spikes based on the APMF of the axon hillock current with our algorithm was tested in case of a 2D array of neurons and a 3D volume of neurons (Fig. 4A). We quantify the performance of the algorithm in reconstructing spike location and time in each case. In both cases, spikes in neurons at each time step, of 500 µs, were generated as a binomial process with a probability $f$ (see Methods for details). In the 2D case, neurons are placed in the plane parallel to the diamond NVC layer, at a spacing of near cell soma size separation of 10 µm, and spikes are assigned as described in methods. Experimental NVMMs are generated by summing of individual NVMMs and adding noise (Gaussian or Shot noise). Figure 4B illustrates an example reconstruction where we observed that most of the spikes are detected and marked correctly in space and time (Fig. 4B, left, 2D case).

The performance of the algorithm for proper detection of spikes in space and time can be obtained with $d' = z(hit\ rate) - z(false\ alarm)$[54]. The algorithm is very robust in the sense that it correctly rejects absence of spikes even in large noise, thus leading to a very high correct rejection rate or very low false alarm. Given the extremely low percentage of false alarms ($<< 1\%$) with the naturalistic sparse firing rates considered, the $d'$ value is very large in most cases even with noise. Hence, to be conservative, we quantify accuracy of reconstruction based on the fraction of correctly marked spikes by total spike instances marked by the algorithm. This would mimic a real situation when actual imaging is performed in the absence of knowledge of all possible spikes that could occur. In the no noise case, the performance was 83.61+/-2.17%, and in the case of added Gaussian noise corresponding to an SNR of -11.8736 dB (see Methods for SNR calculations), the performance was 83.82+/-2.09%. Here, we conclude that matching pursuit algorithm can have inherent errors in reconstruction, even without noise, but the reconstruction shows high resilience to Gaussian noise.

The 3D setup is a more complex case, as NVMM of a single neuron closely resembles, not only neurons in the same lateral plane, but to nearby neurons in multiple directions. Further, due to varying distance from the diamond NVC layer, neurons in different axial plane have varying magnitudes of NVMMs. We show that such an ensemble of neurons can be reliably reconstructed (Figure 4 D, F). We found performances of 68.77% (+/-1.41) without noise and 71.7281% (+/-1.1886) with Gaussian noise. Similar to 2D case, 3D reconstruction is resilient to Gaussian noise at SNR -9.46dB. However, a higher signal-to-noise ratio is needed to perform reconstruction in 3D setup due to more complex correlation structure in the columns of the dictionary. The overall detailed results of performance for the 2D and 3D cases, with and without noise are provided in Table S1.

To compare noise resilience in another method, we performed Moore-Penrose pseudoinverse[55] based reconstruction of the same linear inverse problem in 3D setup (Fig. S1). In pseudoinverse-based reconstruction, without noise reconstruction is 96.19%, but it shows high sensitivity to noise, as the performance drops to 20% percent and a very low fraction of detected spike instances. Due to high correlations in pairs of columns of the dictionary representing closely spaced neurons, the dictionary is an ill-conditioned matrix, and hence pseudoinverse solutions are highly sensitive to noise.

Based on the magnitude of the actual magnetic field, ideally, 3.0 ms, 3.5 ms and 4.0 ms NVMMs should be used as the dictionary elements. The above frames contain the highest magnetic field signatures and are closest to the axon hillock activity. Hence, initially, we performed 2D and 3D reconstruction simulations with these time points. Table S2 contains the details of performance with the above time points (3, 3.5, 4 ms) constituting the dictionary elements. While the 2D reconstruction were decent, the 3D reconstruction was not satisfactory. A critical difference in these set of time frames is the opposite directions of magnetic field signatures between NVMM at 3 ms and 3.5ms (Fig. 2A). We believe that the algorithm, primarily in the 3D case, is sensitive to this directionality of the magnetic field.

*Analysis of resolvability of spatially and temporally close single neuron APs*

The most important issue in reconstruction is resolvability of spikes in nearby neurons (space) and time. The limits of resolvability or resolution of reconstruction would determine if single cell resolution imaging can be performed with the proposed technique and algorithm. We analyse whether only two nearby neurons, separated by distances comparable to single soma size, ~10-20$\mu m$, can be reconstructed by the algorithm. In the entire population of 2D or 3D case, only two nearby neurons are assigned one spike each with a fixed time difference of $\Delta t$. Experimental time series for these two spike events is formed and the algorithm is applied to reconstruct the spike times and location. Initially, no noise is added to experimental time series. Also, the neurons are in same geometric arrangement as shown for performance reconstruction (Fig. 4A).

Single best reconstruction cases are shown for the 2D (Fig 5A. *Left*) and the 3D (Fig. 5A. *Right*) cases. In the 2D case, cells separated by 20$\mu m$ and with spike time differences of $\Delta t$=1ms are correctly reconstructed. Similarly, in a 3D case, an axial pair of neurons separated by 7$\mu m$ and with spike time difference of $\Delta t$=0.5ms are accurately reconstructed.

To better understand resolvability of nearby cells by the algorithm, we successively vary the lateral separation, distance between two parallel neurons in the same plane (2D case), from 200$\mu m$ to10$\mu m$ in steps of 10$\mu m$ with fixed $\Delta t$. These individual 20 experimental time series, with no noise added, form 20 different cases of reconstruction. Two different spiketime differences of $\Delta t$=0.5ms and $\Delta t$=1ms have been studied and represented in Fig. 5B (*left* and r*ight* respectively). For each separation case, the locations of spiking neurons are marked by black vertical ticks. Cases of reconstruction where the spike timings of spike instances marked by the algorithm exactly matches the spike timings of the actual case are shown by green ticks below the corresponding black ticks. Cases of reconstruction, where the algorithm marked an incorrect spike are shown with red ticks. Cases of reconstruction where the algorithm did not mark any spike instance are empty.

For $\Delta t = 0.5\ ms$ (Fig. 5B. *Left*), we observe neurons separated by more than 130$\mu m$ are correctly reconstructed. However, when spiking neurons are brought closer in space, mostly spike instances are not marked by the algorithm or incorrectly marked by the algorithm. Nearby neurons have highly correlated dictionary elements, and experimental timepoint maps where activity from both neurons are present can resemble some other nearby neuron. Therefore, continuous stretches of $p1$ indices are not formed (see Methods-Algorithm) and hence, no spike instance or incorrect spike instances are marked by the algorithm. However, for one millisecond time difference cases (Fig. 5B. Right), we observe correct reconstruction till minimum separation of 20$\mu m$. The lateral resolution in one millisecond case improves due to lesser temporal overlap of DNVMM signals of nearby neurons.

Further, we test the extent to which Gaussian noise can affect the above reconstruction cases of nearby neurons. The above reconstruction of two nearby neurons becomes stochastic by addition of Gaussian noise to the experimental time series. Therefore, a minimum SNR is required for the algorithm to perform correct reconstructions and a lower minimum SNR is an indicator of high resilience of reconstruction to Gaussian noise.

To estimate minimum SNR for the 2D (lateral separation) and 3D (axial separation) case, we perform multiple repetitions of reconstruction of two nearby neurons, as above (Fig. 5) but with additive Gaussian noise. To consider the worst-case scenario, the accuracy of reconstruction has been studied in 2D case, for one spike each from two nearby neurons which are laterally separated by 10 $\mu$ms and $\Delta t = 0.5\ ms$. For the 3D case, one spike each from two nearby neurons which are axially separated by 7 $\mu m$ and $\Delta t = 0.5\ ms$. By varying levels of Gaussian noise factor, we vary signal-to-noise ratio and simulate reconstruction (run algorithm) for 50 independent repetitions of the two-neuron case for both 2D and 3D. The minimum SNR is considered as a point where the standard deviation of reconstruction drops to zero (for these 50 repetitions). For each repetition, a correct reconstruction, where algorithm marks correctly the neuron and its spike time both, is given value 1 and 0 is given otherwise.

We observe decreasing standard deviation and increasing mean of correct classification percentage (average of individual 0/1 values) to 1 as the Gaussian noise factor decreases for 2D case (Fig. 6A *Left*). Similar trends are observed for the 3D case as well (Fig. 6A *Right*). The point of minimum SNR is marked as a dashed vertical line for both the 2D and 3D cases in Fig. 6A. Since the Gaussian noise factor is not linearly related to SNR (see Methods-SNR calculations), the actual mapping between them is shown Fig. 6B (2D, *Left* and 3D, *Right*). We find -13.9 dB minimum SNR for 2D case (at Gaussian noise factor=0.0061) and -10.2 dB minimum SNR (at Gaussian noise factor=0.0046) for the 3D case. A higher minimum SNR required for the axial case is due to more complex correlations of any single neuronal DNVMM to other neuronal DNVMMs in nearby volume as compared to lesser number of nearby correlated neuronal DNVMMs in a plane in the lateral case. We observed a clear certainty in individual events of reconstruction (for 50 repetitions) when the SNR is higher than the minimum SNR for both the lateral (Fig 6C *Left*) and axial case (Fig 6C *Right*).

A negative minimum SNR shows high resilience of the proposed algorithm to Gaussian noise. Further, it implies that reconstruction might be possible in diamond NVC experiments with lower magnetic field sensitivity, where the noisy magnetic field data can be compensated with prior information based on the axon-hillock's APMF signature in the dictionary.

While the experimental maps are expected to carry Gaussian noise, we also evaluate the above minimum SNR for shot noise dominated experimental maps. All analysis remains the same, but instead of Gaussian noise, shot noise is added to the experimental maps (see Methods). Fig S2 shows the results of the same analyses as in Fig. 6. for reconstruction in shot noise based experimental maps. Similar trends are observed for shot noise analysis. However, the minimum SNR required for shot noise maps is found to be significantly higher as compared to gaussian noise maps. We find 14.08 dB minimum SNR for 2D case (at shot noise factor=0.2001) and 20.18 dB minimum SNR (at shot noise factor=0.1001) for 3D case. The shot noise maps contain noise proportional to signal magnitude, having more jitter at high magnetic field values. Therefore, shot noise affects the overall features in the DNVMMs and hence, requires more SNR to be able to reconstruct accurately. Further, the residual noise, if any, in shot noise maps are highly correlated to, in terms of features, to the neuron DNVMM that was subtracted from the experimental time series. Therefore, it does not lead to complete removal of a neuron's signature from experimental time series, even when a spike for that neuron has been assigned. This effect, further, imposes high SNR requirements for shot noise map reconstruction.

**Discussion**

In this work, we first show that APMF of the axon hillock in mammalian neurons can serve as a specific signature for detecting spiking activity in single neurons in a 3D volume of brain tissue. Previously reported imaging APMF of entire worm axon or reconstructing current carrying wires[27,30,39,56] suggests the usage of entire axonal APMFs, which would render 3D magnetometry based imaging near impossible because of routes and long lengths of axons in brain tissue. We have estimated magnitude of mammalian pyramidal neuron APMF for the axon-hillock segment to be 36pT, which is 2 orders of magnitude larger than other locations of a neuron. Such specific APMF signatures thus allows reconstruction of single neuron activity in a 3D tissue. Based on recent experiments and theoretical advancements, we believe that sensitivities of widefield magnetic diamond NVC imager will be within reach to image APMF signals. We have developed and presented an algorithm to find neuronal spike timing and location from 2D NVMMs. We show it is possible to perform spike activity reconstruction of hundreds to thousands of neurons located in a 2D layer or 3D volume. We also show the spatiotemporal limits of correct reconstruction to be in line with near simultaneous firing of spikes at single cell spatial resolution, provided sufficient sensitivity in the experiment.

We highlight the Gaussian noise resilience of the algorithm proposed. In cases of Gaussian noise, where minimum SNR required is low in the range of -10dB to -20dB, an experimental setup with sensitivity nearly equal to peak magnitude of APMF will be sufficient to reconstruct neuronal spiking activity. Therefore, in widefield diamond NVC experiments, where spatial resolution, temporal resolution and volume normalized sensitivity are tightly coupled, application of the proposed algorithm on larger pixels might allow same reconstruction accuracy but an increase in temporal resolution or higher sensitivity experiments.

However, in a shot noise dominated regime, information in signal patterns are significantly lost due to addition of noise correlated to signal magnitude. In this regime, signal to noise amplitude ratio should nearly be 10, to achieve a decent reconstruction. This SNR requirement would demand a sub-picotesla DC diamond NVC magnetometry, which has not been demonstrated yet. However new techniques, with genetic manipulations for specific enhancement of the axon hillock associated APMF or converting the axon hillock associated current to an AC magnetic field, which can be detected with AC magnetometry where much lower levels of detection is possible[57–60], would allow single cell resolution mapping of APMFs with currently available magnetometry techniques.

The AP magnetic field signal is approximately 2 ms in timescale, which will fall in DC signal range as compared to diamond NV measurement protocol, which can span ~10-2000 μs in time depending upon the coherence time $T_2$ on the sensor[61,62]. As demonstrated in the later sections on reconstruction of spiking activity, we suggest that DC vector magnetometry at, at least ~ 1 pT/μm$^{3/2}$/ $\sqrt{Hz}$ will be required to reach close to single cell resolution spike detection, with our developed algorithm. In Barry et al PNAS 2016, the authors demonstrated 34 nT μm$^{3/2}$/ $\sqrt{Hz}$ DC field sensitivity and they expect a 100-fold improvement with engineered diamond, Ramsey protocol and optimized collection methods. Additionally, recent work[56,58,63,64] suggests use of advanced quantum manipulation methods to reach closer to the quantum projection shot noise limited DC magnetometry. For example, Liu et al[65] demonstrated Ancilla assisted sensing to increase DC field sensitivity and also performed a rejection of 1/f low frequency noise in DC magnetic field measurements. Sub-picotesla magnetometry has been already been demonstrated for AC field sensing[57]. Various expected DC field magnetometry[62] related research expect the ensemble diamond NVC sensitivity to reach pico-tesla levels, where we expect to see at least, blurred single neuronal APMF signatures.

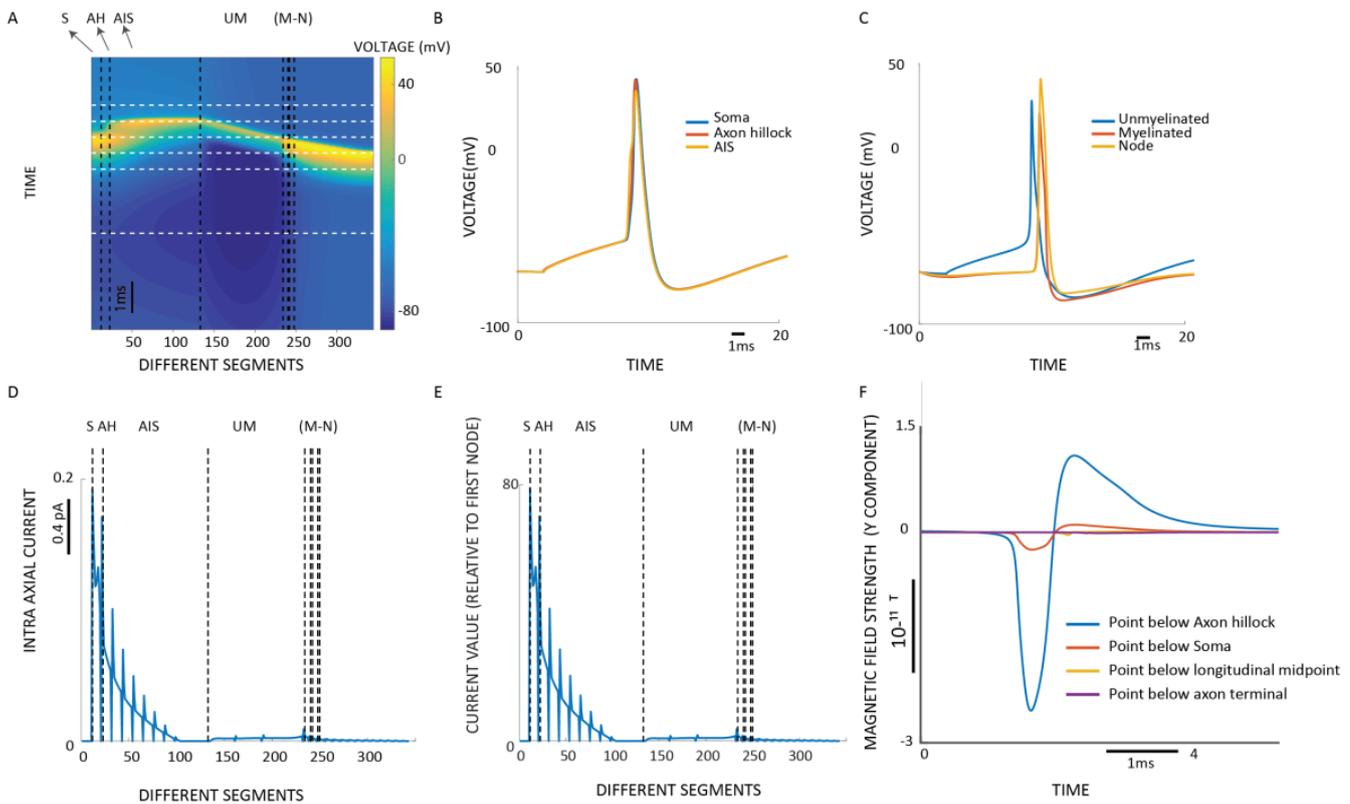

**Figure 1:** Simulations of membrane potential and estimation of intra-axonal currents. **(A)** Color map of membrane potential as a function of time and spatial segments of the neuron. Dashed vertical lines (black) mark boundaries of different neuronal regions on the map. Alphabet codes denote names of neuronal segments. S-Soma, AH-Axon hillock, AIS-axon initial segment, UM-unmyelinated region, M-N along with repeating vertical lines marks repeating regions of (myelinated axon – node of Ranvier)$_n$ . Only first two myelin node boundaries are marked by vertical lines for clarity in figure. The Action potential (AP) originates in the most distal end of AIS region and propagates in a bi-directional manner. Horizontal white lines denotes time points considered in further analysis (Fig 2) **(B)** Membrane potential profile of particular segments of Soma (Blue), Axon Hillock (Red) and AIS(Yellow). **(C)** Other membrane potential profiles of particular segments of Unmyelinated axon (Blue), Myelinated region (Red) and Node of Ranvier (Yellow). **(D)** Peak intra-axonal current profile of different neuronal segments. Maximum Intra-axonal current flows through most proximal segment of axon hillock. **(E)** Peak Intra-axonal current profile normalized by peak intra-axonal current in second node of Ranvier. Peak currents in axon hillock are nearly two order larger than other axonal regions. **(F)** Y component of magnetic field strength measured at a point perpendicularly below different points on the neuron has been shown. Measurement point below Axon hillock (blue), Soma(red), longitudinal midpoint (yellow), axon terminal(purple) have been compared.

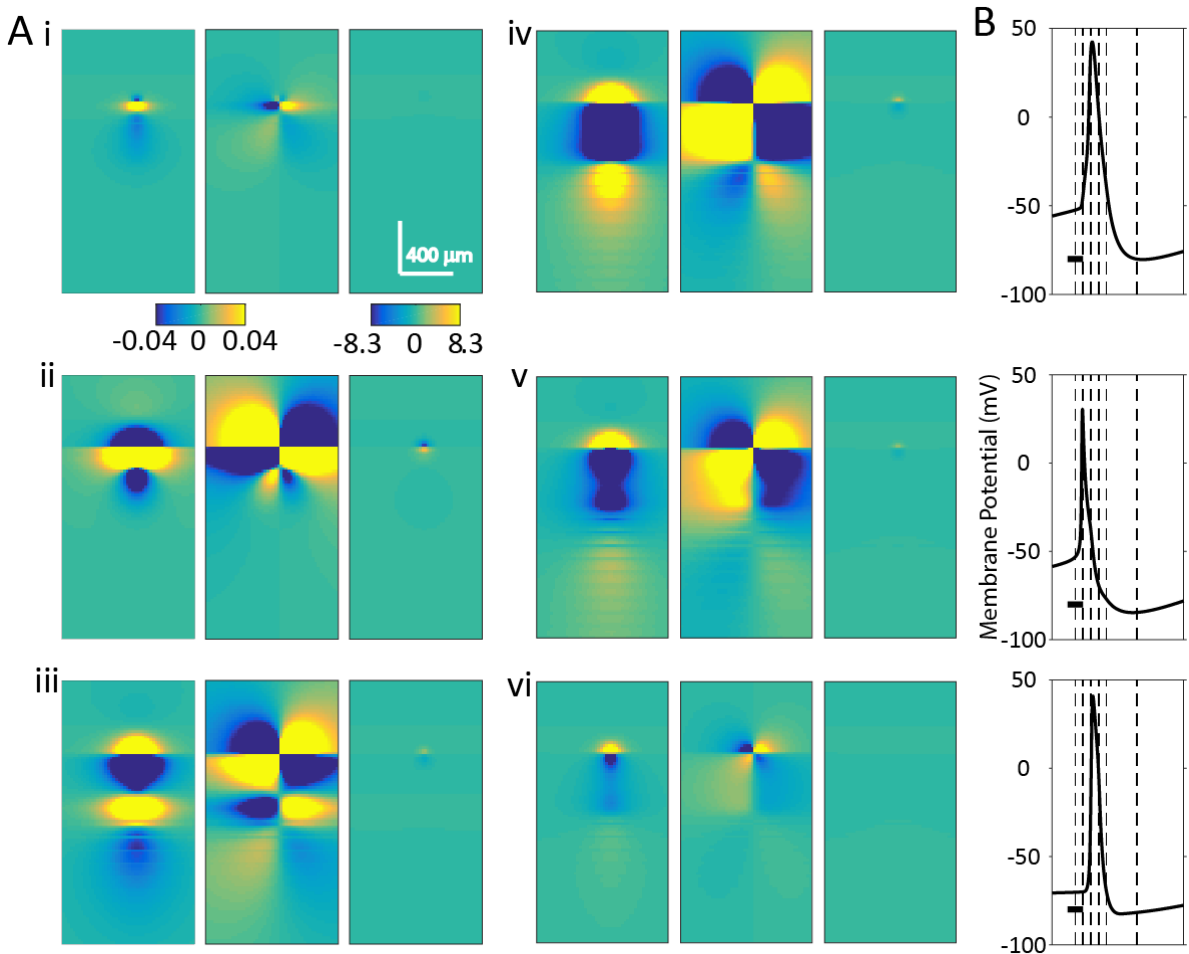

**Figure 2:** 2-Dimensional Diamond Nitrogen Vacancy vector Magnetometric maps (DNVMM) of a single mammalian spike. Color map of AP propagation in a cortical pyramidal neuron (Fig 1), is marked with horizontal white lines marked i, ii, iii, iv, v, vi to signify different time points during AP propagation. **(A)** 2D DNVMMs corresponding to different timepoints are shown in this panel. Markings – i-2.5ms, ii-3ms, iii-3.5ms, iv-4ms, v-4.5ms, vi-6.5ms. First two columns show intentionally reduced color axis of Y and Z component of magnetic field in picoTesla. Color axis is saturated to show patterns in maps, where magnetic field is approximately two orders lower in magnitude. The third column contains maps for Y component of magnetic field with color axis scaled to maximum magnitude of magnetic field across all timepoints, 8.3pT in this DNVMM. In the third column, APMF signatures are visible only in timeframes corresponding to AP passing thorough axon hillock region. **(B)** Corresponding mapping of different timepoint 2d maps to membrane potential of neuronal segments. Axon hillock proximal end (top), axon initial segment(middle) and first node of Ranvier(bottom) have been shown with timepoints marked by vertical lines corresponding to timepoint i, ii, iii, iv, v, vi sequentially. Scale bars in all three plots correspond to 1-ms.

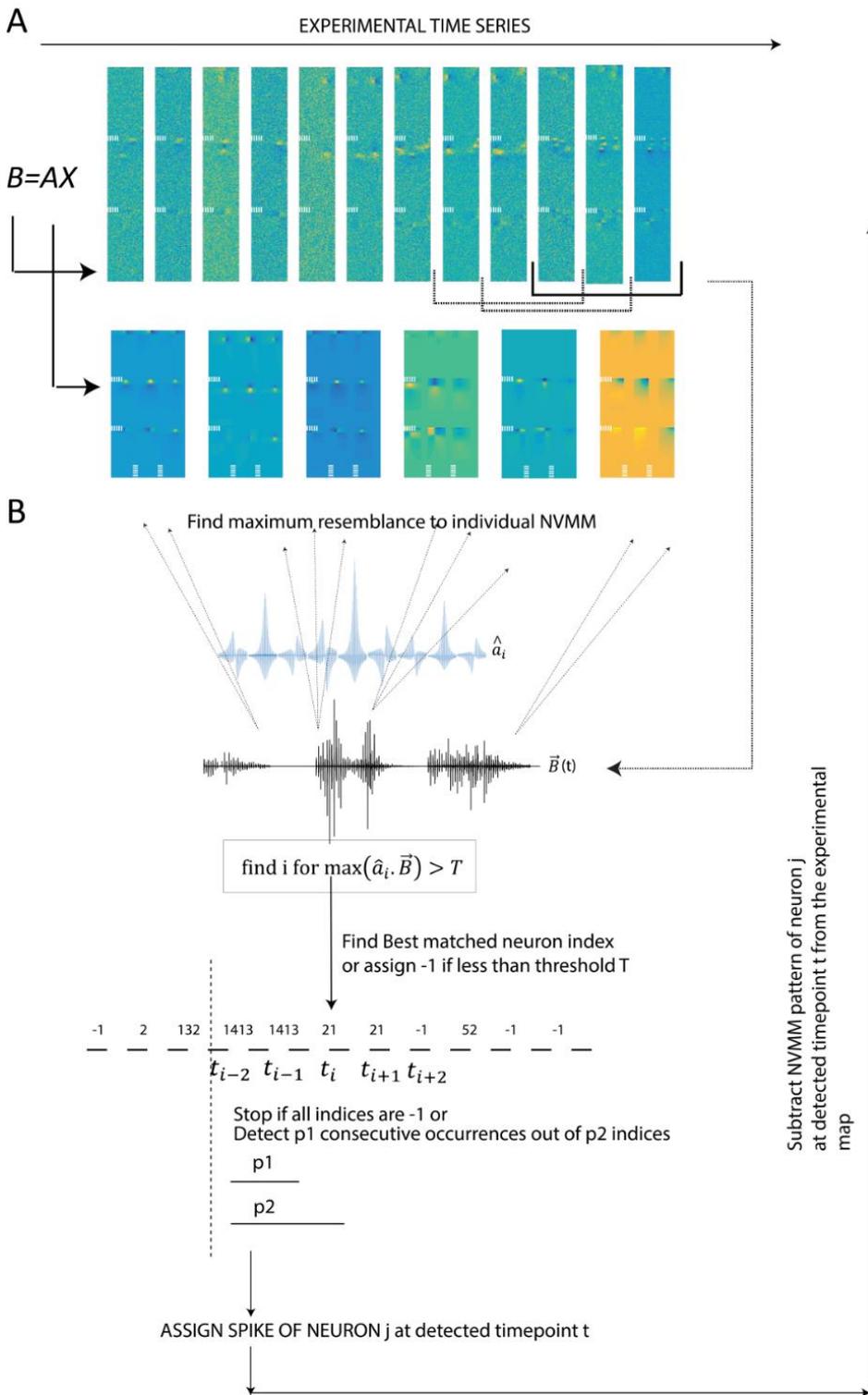

**Figure 3:** Schematic of the reconstruction algorithm for detecting spike timing and location from 2D NVMMs. The algorithm is recursive in nature. In each step, a set of experimental time series maps are given as input. (**A**) An example experimental time series maps (top) and example dictionary maps (bottom) are shown. Three successive time points of the time series are concatenated (three such successive groups are marked at the end) to be use further. At each time point, the experimental 2D map consists of 3 maps corresponding to X, Y and Z components of the magnetic field. Dictionary elements are also set up accordingly with three time points of the AP. (**B**) All 2D maps and dictionary elements are converted to 1D signals, shown as 1D vectors in the schematic. Each time segment is matched with all dictionary elements and resemblance is determined by a dot product. Series of best neuron indices are processed further to look for consecutive occurrences of same neurons and transitions from one neuron to the next. Based on best neuron indices, DNVMM signals of selected spike instances (space and time of spike) are subtracted from the experimental maps and residuals are equated to experimental maps for next iterations. The algorithm stops when no APMF signature is detected from experimental maps.

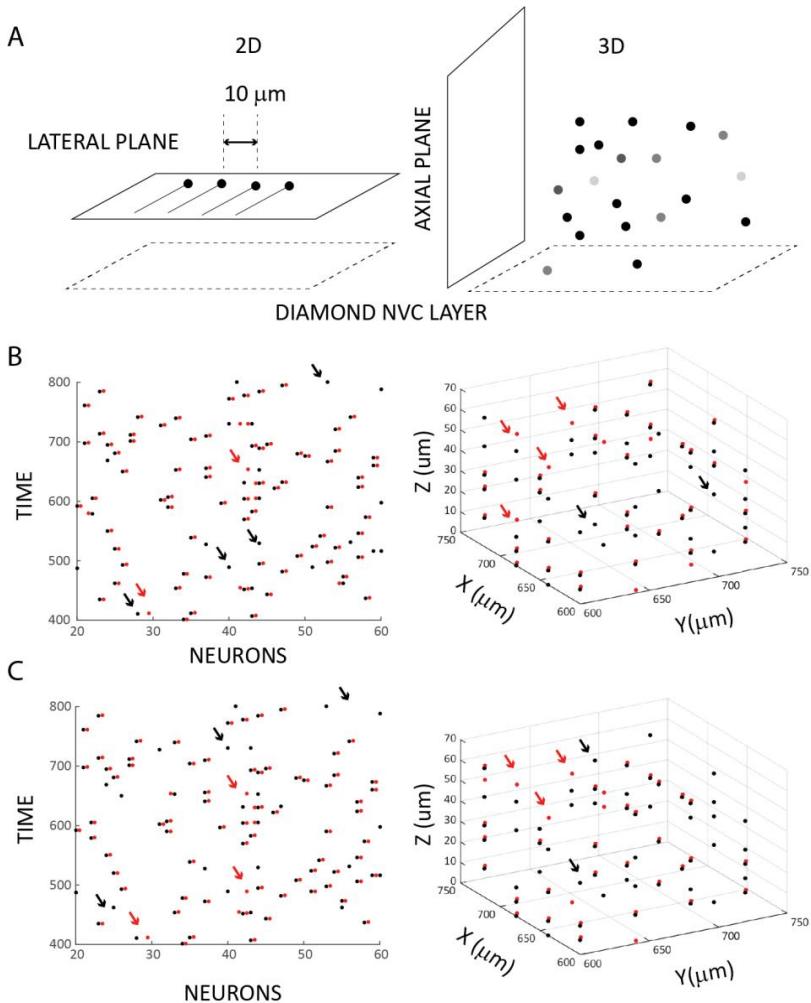

**Figure 4:** Population reconstruction performance in 2D layer and 3D volume. **(A)** Schematic showing arrangement of neurons in 2D layer in a plane parallel to diamond NVC layer (left) and that of neurons in a 3D volume placed above diamond NVC layer (right). **(B)** Visualization of performance of the algorithm, for a 2D case without noise (left), where black dots are actual spike instances and red dot show spike instances marked by the algorithm. X axis denotes the lateral position of the 2D neuron and Y axis denotes time. Only, few neurons are shown in this figure for clarity. For correct reconstruction, red and black dots overlap but have been slightly shifted to compare reconstruction. Few false positives are marked with red arrows and missing correct cases have been marked by black arrows. The same for a 3D case without noise is shown to the right. Representation of spike instances, correct detection, false positives and misses are same as to the left. All actual/reconstructed events have been integrated in time into one 3D plot, where the three dimensions correspond to cell soma location of neurons. **(C)** Same as **(B)**, with added Gaussian noise.

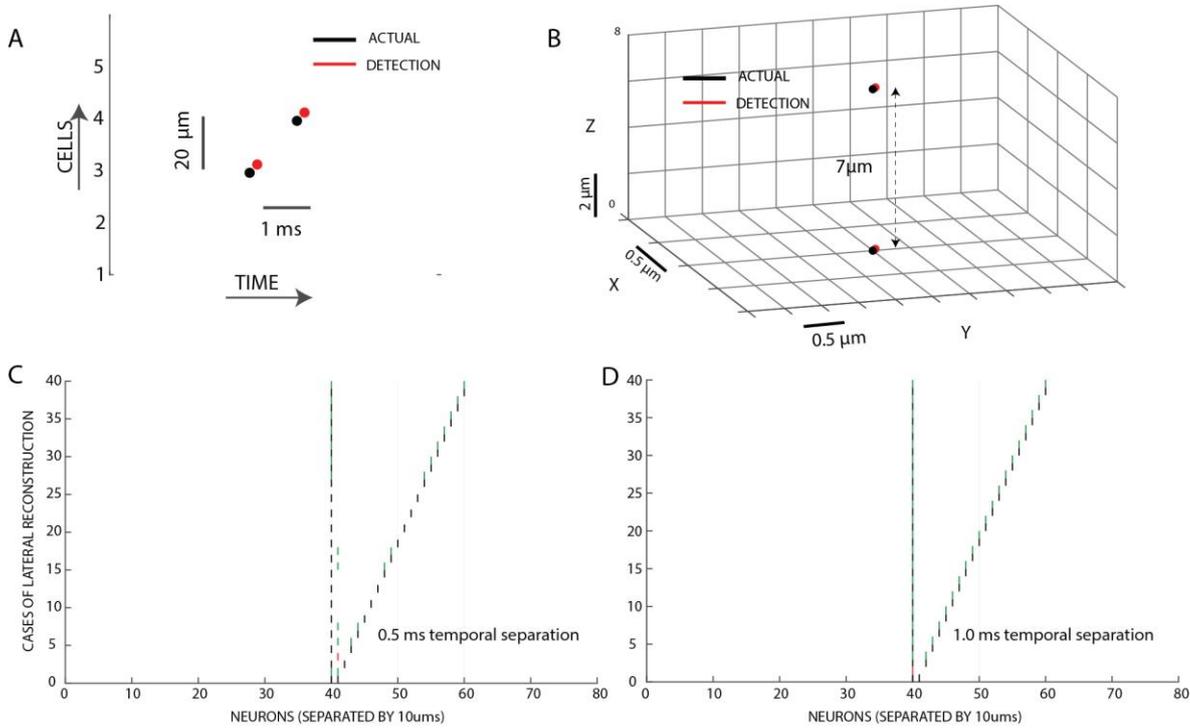

**Figure 5:** Single-cell spatial resolution by reconstruction from Gaussian-noise dominated 2D NVMMs. **(A)** *Left* Plot showing two laterally separated neurons (both neurons located in the plane parallel to diamond NVC layer) separated by 20μm, firing near simultaneously, one millisecond difference. Actual Spikes timings and location have been marked with black dots. Reconstruction of spike instances (timing and location) from the proposed algorithm have been marked by red dots. The dots coincide but have been slightly shifted for visualization. **(A)** *Right* Plot showing two axially separated neurons (neurons separated along plane perpendicular to diamond NVC layer) separated by $7\mu m$ and $0.5ms$ spike time difference. Color code same as Fig. 5 (A) *Left* **(B)** *Left* Plot showing case by case reconstruction accuracy (20casesX2, actual and algorithm) with varying lateral separation (per neuron number, $10\mu m$ shift) between two neurons located in the same diamond NVC plane. Spike time difference fixed at $0.5ms$. Black vertical lines shows location of actual spiking neurons. Green vertical lines show spike instances marked by algorithm, if spike times are detected correctly. Red vertical lines show spike instances marked by algorithm, if spike times are incorrect. **(B)** *Right* Plot showing case by case reconstruction accuracy (20casesX2, actual and algorithm) with varying lateral separation (per neuron number,$10\mu m$ shift) between two neurons located in the same diamond NVC plane. Actual Spike time difference fixed at $1ms$. Color code and Y axis same as of Fig 5 (B) *Left*

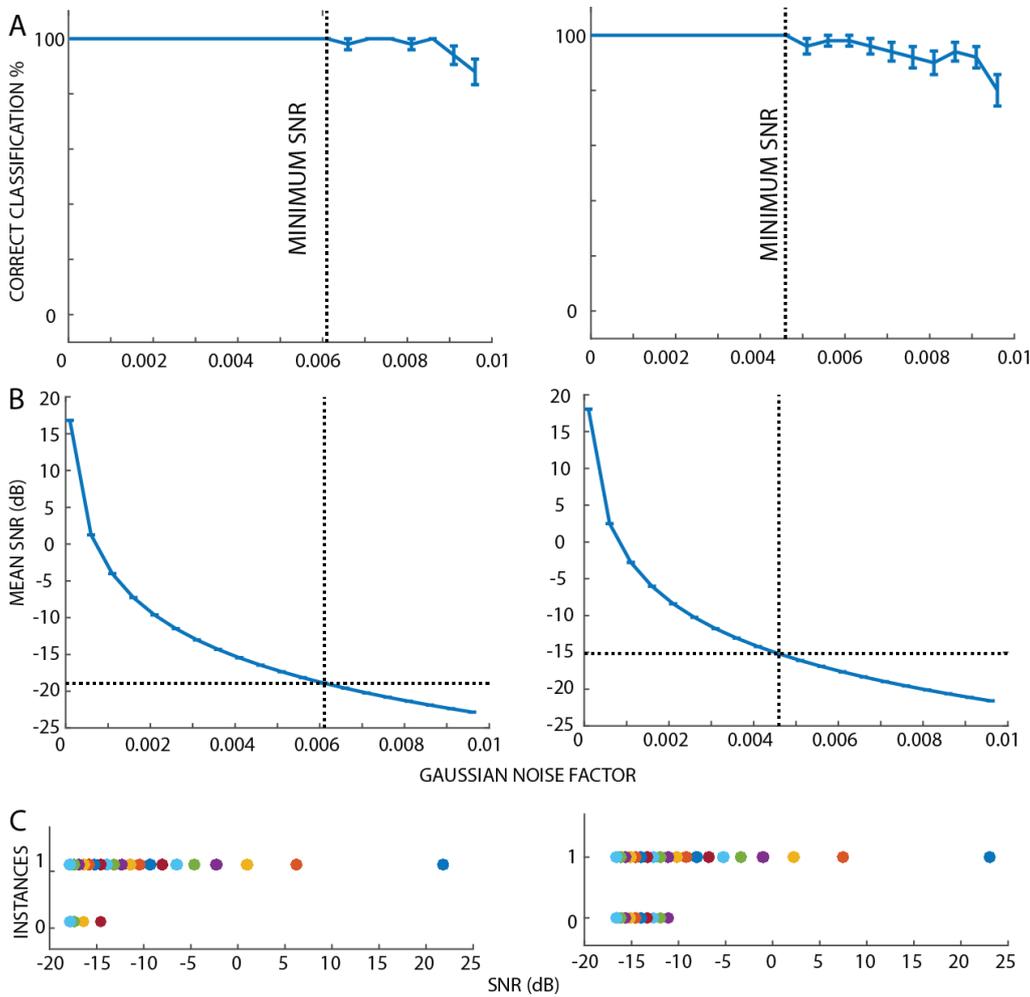

**Figure 6:** Minimum required SNR (Min. SNR) requirements for lateral (*Left panel*) and axial (*Right panel*) separation case in Gaussian-noise DNVMM maps. **(A)** *Left* Plot of correct classification percentage versus gaussian noise factor (see Methods, SNR Calculations) for lateral separation case (two neurons ,10$\mu m$ apart, spike time difference 0.5$ms$). Min. SNR has been marked as the point where the standard deviation of correct classification percentage drops to zero (in 50 repetitions) **(A)** *Right* Plot of current classification percentage versus gaussian noise factor for axial separation case (two neurons, 7$\mu m$ apart, spike time difference 0.5$ms$). Vertical line marks point of Min. SNR. **(B)** *Left* Plot of mean SNR (dB) versus gaussian noise factor for lateral separation case. **(B)** *Right* Plot of mean SNR (dB) versus gaussian noise factor for axial separation case. (A) and (B) plots share same X axis title– Gaussian noise factor. **(C)** *Left* Example individual instances of correct(1)/ incorrect(0) reconstruction by proposed algorithm versus SNR(dB) for lateral separation case. A clear shift in certainty of reconstruction is observed at SNR above Min. SNR. **(C)** *Right* Example individual instances of correct(1)/ incorrect(0) reconstruction by proposed algorithm versus SNR(dB) for axial separation case. Left and Right panels of (C) share same X axis title – SNR. Different colors are used to distinguish different individual instances.

**Methods**

*Simulations of intra-axonal currents in a cortical pyramidal neuron*

A realistic neuronal model[34] was implemented in NEURON and used to simulate the membrane potential. Here, we briefly describe the simulation of membrane potentials. The cable equation that governs membrane potential is –

$$\frac{1}{2\pi a(r_i+r_o)}\frac{\partial^2 V_m(z,t)}{\partial z^2} = C_m\frac{\partial V_m(z,t)}{\partial t} + J_{ion} - J_m \qquad (1)$$

The discretized version of the equation was solved by implicit PDE solvers in NEURON. The temporal resolution of the voltage and current values available after the NEURON simulations was 10 $\mu s$. These data were imported and analysed further in MATLAB (Mathworks) with custom written routines.

The pyramidal neuron used in the model has the following types of compartments – cell soma, axon hillock, action initial segment, unmyelinated region and repeated regions of myelinated axon and node of Ranvier.

Intra-axonal current flowing across each segment of the neuron at a given time instant was calculated from the following equation –

$$i_{compartment,i}(t) = \frac{V_m(i,t)-V_m(i-1,t)}{r_i\frac{dl}{\pi a^2}} \qquad (2)$$

Where $i, i-1$ denote the adjacent segments of the neuron. $r_i, dl, a$ are resistivity, length and radius of the segment respectively.

*Calculation of magnitude of APMF and simulations of experimental 2-D DNVMM*

By application of Biot-Savart's law, we calculated magnetic field $\vec{B}$ at a measurement point $\vec{r}$ by summing over different segments $n$ of the pyramidal neuron.

$$\vec{B}(\vec{r},t) = \sum_j^N k\, i_j(\vec{r_j},t)\frac{\vec{dl_j}\; X\;(\vec{r_j}-\vec{r})}{|\vec{r_j}-\vec{r}|^3}, \text{ where } \vec{B}=[B_x\; B_y\; B_z]^T \qquad (3)$$

Where $\vec{dl_j}$ is a length vector along the segment, $\vec{r_j}$ is the position vector of the segment $j$, $\vec{r}$ is the measurement point, $i_j(\vec{r_j},t)$ is the current in the segment, $k = \mu_o/4\pi$, and the summation is over all segments $j = 1,2,3\ldots N$ of the pyramidal neuron.

In order to estimate the magnitude of the mammalian APMF, a measurement point was selected perpendicularly below different segments of the pyramidal neuron at a standoff distance $d$ from the centre of the compartment. The magnetic field at the measurement point was calculated by Eqn. 3, above. APMF magnitudes at 4 different measurement points shown in Fig. 1F were selected as follows: perpendicularly below cell soma, below Axon hillock, below longitudinal midpoint of the pyramidal neuron and below the axon terminal end.

NVMMs are comprised of magnetic field values at multiple 2D spatial points calculated by varying the $r$ vector (Eqn. 3, above) at different points in the diamond NVC plane. NVMMs are 50 pixels X 100s pixel in size, with each pixel size equal to 20 $\mu m$ X 20 $\mu m$. A time series of NVMMs are obtained by simulating the NVMM at different time points during AP propagation in the pyramidal neuron.

*Mathematical formulation of the inverse problem and the reconstruction algorithm*

We solve for the inverse problem in Eqn. ii (Main Text) where $B$ is the experimentally acquired 2D DNVMM frames (with Gaussian noise or shot noise, see below in the section on Generation of spikes and time series of maps).

$p_x = 100$ number of pixels of the NV sensor in the $x$ direction

$p_y = 200$ number of pixels of the NV sensor in the $y$ direction

$n$ number of neurons in 2D plane or 3D volume

$n_{tp}$ number of AP timepoints considered in the reconstruction (set to 3)

A dictionary matrix , size $3n_{tp}p_xp_y$ X $n$ (factor of 3 for $B_x$, $B_y$ and $B_z$ components)

$A_i$ columns of the dictionary with the $3n_{tp}p_xp_y$ elements after concatenation of each component of $n_{tp}$ time points

$B_t$ Concatenated Experimental map at timepoint $t, t-1, t-2 \ldots t-n_{tp}+1$ , vector of length $n_{tp}p_xp_y$

$T$ is the threshold of projection value for detection of spiking. Threshold values of $10^{-12}$ and $10^{-11}$ were used for 3D and 2D cases respectively.

$I_B$ is a vector indicating the best matched neuron index in every cycle of the algorithm, with values ranging from 1 to $n$ or -1 (no match) corresponding to each time point $t$

$p2$ total number of successive spike time event scan length for $I_B$ indices set to 3

$p1$ minimum number of occurrences of a neuron required in successive $p2$ elements of the $I_B$ vector to be considered as a spike, set to 2

$p$ running value of $p1$

Each column of $A_i$ is set to $|\phi_j\rangle$ , the NVMM of a particular neuron. This vector contains concatenated frames of multiple time instances (mainly corresponding to axon hillock activity) and multiple directions ($B_x$ $B_y$ $B_z$). Let $t_1$ $t_2$ $t_3$ ... ... ... ... ... $t_n$ be time instances of sampling. At each time t, we find resemblance of experimental map to columns of the dictionary and assign index of best matched neuron to that time instant.

Part 1: Find $I_B$

At each experimental time point $t$ , find which neuron $I_B(t)$ out of all neurons $n$ resembles most to the experiment map $B_t$

Also, the value of the projection on normalized dictionary elements must be greater than a certain threshold ($T$) for the neuron to be selected or else we place -1 at $I_B(t)$ if $\max\left(\widehat{A_i}.B_t\right) \geq T$ , then $I_B(t) = k = \underset{i}{argmax}\left(\widehat{A_i}.B_t\right)$

else if $\max\left(\widehat{A_i}.B_t\right) < T$ , then $I_B(t) = -1$

if all elements of $I_B$=-1, algorithm stops

If we find $p = p1$ occurrences of a matched neuron ($k$) in a continuous stretch of $p2$ elements on $I_B$, a spike of the neuron $k$ is detected. Precise timing of the neuron $k$ spike is determined by another search for time instant where subtracting $k$ neuron's signal leads to maximum reduction in signal $B$. After subtracting $A_k$ from the signal $B$ for appropriate time points, $B - A_k$, the residual is carried over as signal $B$ for the next iteration. Hence, we detect and subtract signatures of all spiking neurons one by one, at best timing location, until no further detection can be done and norm of signal $||B||$ at each time instant is less than threshold $T$. Notably, there are three parameters namely $T$ , $p1$ and $p2$ that control the output of the algorithm. In particular, we initialize $p$ as $p2$ and subsequently:

Search while $p < p1$ for different $t$

find consecutive occurrences of neuron $i$ in the vector $I_B$ for $p$ times out of moving scan window of length $p2$

On detection of a spike of neuron $k$

find max ($\widehat{A_k}.B_{t0}$ , where $t0$ ranges from $t - n_{tp}$ to $t + n_{tp} - 1$ & only where $I_B$ equals $k$)

Choose the argument of this maximum as spike timing $t_{spike}$ of neuron $k$ followed by subtraction as mentioned before.
It is to be noted, after subtracting DNVMM corresponding to a particular spike of a neuron, when we go to recalculation of $I_B$ from the new experimental map, we do not evaluate $I_B$ indices over all time points of the experiment. We revaluate $I_B$ only within the time points which have changed due to subtraction of DNVMM of the last spiking neuron. The process is continued by,

end for a particular $p$ and subsequently

decrement $p=p$-1

go to $search\ while\ p < p1$

*Running algorithm*

The algorithm was quantified in two different cases – 2D case, where the neurons are located in a plane parallel to the diamond NVC layer and a 3D case, where the neurons were distributed in 3D volume, randomly oriented, mounted over a diamond NVC layer. In the 2D case the dictionary matrix is comprised of individual NVMMs of 80 neurons laterally shifted by 20 microns. In the 3D case there are total 6250 different NVMMs from randomly oriented neurons in the 3D volume of $1mmX2mmX\ 70\mu m$. The placement of cell somas/axon hillocks was done in grid like manner by placing 25X25 neurons in each plane parallel to diamond NVC layer, and stacks of 10 such planes with varying perpendicular distance, z coordinate, from the diamond NVC layer. The spatial resolution of this grid was $40\mu mX40\mu mX7\mu m$ in X, Y and Z respectively. After placement of cell soma, the direction of the neuron was randomly chosen from 10 different orientation angles between 0 to 90 degrees. The corresponding NVMMs were added to the 3D case dictionary.

The experimental map was constructed as a convolution of spike timing vector and individual NVMMs time series. The timing resolution was kept at 0.5 ms and total simulation time was set to 600ms. Spike timing was assigned by method specified above.

*Generation of neuronal spikes and time series of maps*

The probability of spike of a neuron at a time instance is given by $f$, a factor that controls the spatial and temporal density of firing. Higher $f$ will lead to more spatially and temporally sparse firing. For each neuron, $f$ is a binomial probability. At each time instant, we generate a uniform random number $r_i$ ranging between 0 and 1 for each neuron $i$. A spike occurs in neuron $i$, if $r_i > f$. Thus the spike times of each neuron is independent of each other. After assigning spikes by the above stated method, for each neuron, to simulate a refractory period [43,66], a neuron is not allowed to spike for a period of 5ms following a spike. The factor $f$ for the 2D case was adjusted to be 0.994 and for the 3D case to 0.9996 so that sparse firing in the population, as in cortex is observed[67].

For 3D performance, some additional spike times, other than multiple spikes within the refractory period, were removed before applying the algorithm. Spikes of Neurons of folllowing two types in the 6250 element 3D Dictionary (see Methods - Running Algorithm) were removed, and hence these neurons produce no spikes .

Type1- A neuron whose RMS value of 1D vector (concatenated 2DNVM) element in the dictionary is less than the threshold (1pT for 3D case). If we assign spikes to these neurons, they are always rejected in threshold step of the first iteration itself, while running the proposed algorithm. Hence, spikes of these neurons are pre-removed before running the algorithm.

Type2- 2DNVMs were generated by summing neuronal intra-axonal currents in Bio-Savart expression for each z plane and a random orientation (see methods - Running Algorithm). However, for each Z plane and orientation angle, the XY grid was simulated by translation of the map along x and y axis. In this translation, some neurons can be shifted to an extent that the axon hillock related signatures do not fall directly above the diamond NVC layer. Hence, these neurons lack important axon hillock patterns in their DNVMs and are significantly less in their RMS values (in agreement with high axon hillock contribution). Spikes of these neurons are removed, as they are never detected by the algorithm. In this dataset of 3D dictionary generation, these neurons are approximately 40 percent in number.

Type1 and Type 2 have large number of common neurons, whose axon hillock segments are displaced off the NVC layer. Only their long axonal parts, fall perpendicularly over diamond NVC layer.

However, their individual 2DNVMs of both types, are always present in the dictionary during the run of proposed algorithm.

### *Signal to noise ratio (SNR) Calculations*

We add Gaussian or shot noise to experimental time series NV maps in the following manner:
$S = [\, B_t \; B_{t+1} \; B_{t+2} \; \dots \;]$ is the concatenated 1D vector of all 1D $B_t$ experimental maps at different timepoints
$S^{noise}$ is the 1D vector with noise added to each element of vector $S$
$\eta$ is Gaussian or shot noise factor
$randn$ MATLAB function was used to generate a Gaussian random variable with zero mean and standard deviation 1
$rms(S)$ root mean square of all elements of vector $S$
For Gaussian noise, each element of $S^{noise}$ is given by
$$S_i^{noise} = S_i + \eta * rms(S) * randn$$
For shot noise, each element of $S^{noise}$ is given by
$$S_i^{noise} = S_i + \eta * |S_i| * randn$$
Signal-to-noise ratio is given by
$$SNR = 20 * \log_{10} \frac{rms(S)}{rms(S^{noise} - S)}$$
To be noted, shot noise maps are dependent on per pixel magnitude, $|S_i|$ and hence, perturb the experimental maps more at pixels where magnetic field is high. However, Gaussian noise maps get the same standard deviation noise added depending on the term $rms(S)$ , which remains constant for different elements $S_i^{noise}$ of vector $S^{noise}$. Also, noise factor values for lateral and axial cases in experimental maps can't be directly compared, but their SNR values can be compared.

**Author contributions:**
MP, KS and SB conceived the project, SB and KS supervised the work, MP did all simulations and analyses and MP and SB wrote the manuscript in discussion with KS.

Authors declare no competing financial interest

**Acknowledgements:**
 MP thanks MHRD for Institute Fellowship and PMRF, KS acknowledges support from IITB-IRCC Seed grant number 17IRCCSG009, DST Inspire Faculty Fellowship - DST/INSPIRE/04/2016/002284 and AOARD R&D Grant No. FA2386-19-1-4042. SB thanks MHRD and IIT Kharagpur for Challenge Grant and India Alliance for Intermediate Fellowship funding Grant No. IA/I/11/2500270.

**Supplementary Information**

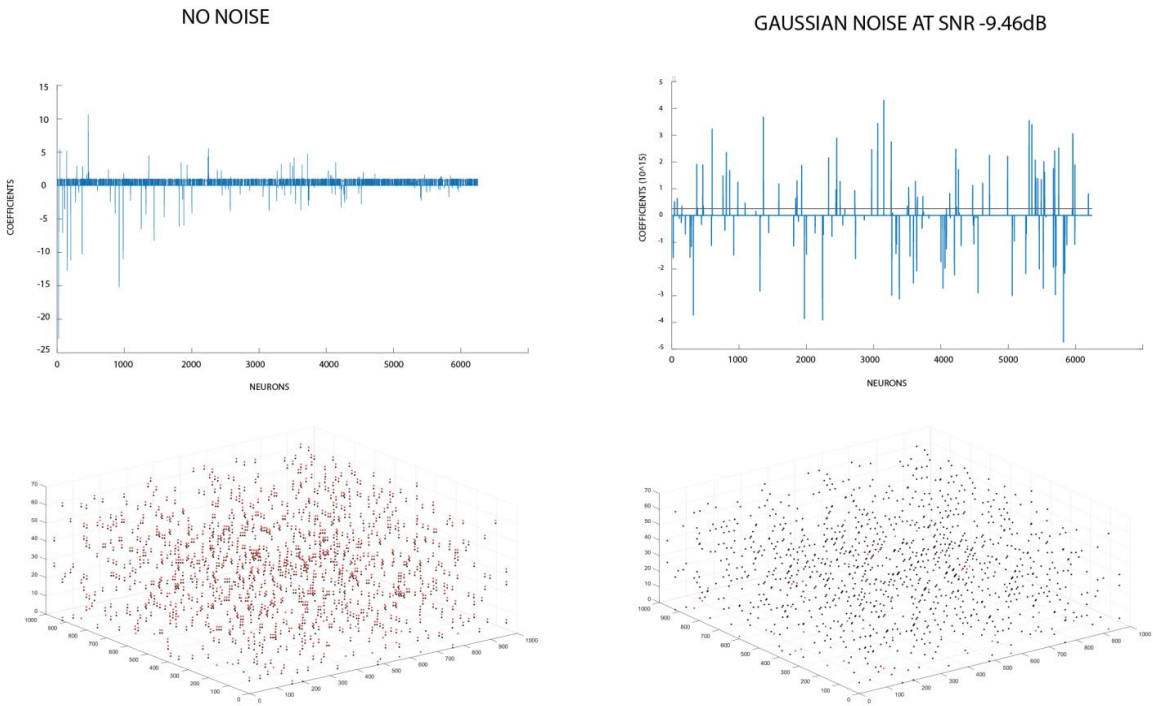

**Figure S1:** A pseudoinverse-based reconstruction of simulated 3D cortical activity. **Top row:** (***Left***) Coefficients after pseudoinverse calculations in no noise case. The horizontal line shows threshold level for considering a spike instance. (***Right***) Similar to figure on the left, but with -9 dB Gaussian noise added to experimental NVMM vector B. **Bottom Row:** (***Left***) 3-Dimensional reconstruction, in no noise case. Black dots represent actual spike instances and red dots represent spike instances obtained from pseudoinverse reconstruction. Without noise reconstruction is nearly 100% percent accurate. (***Right***) Similar to figure on the left, but with noise added to experimental NVMM.

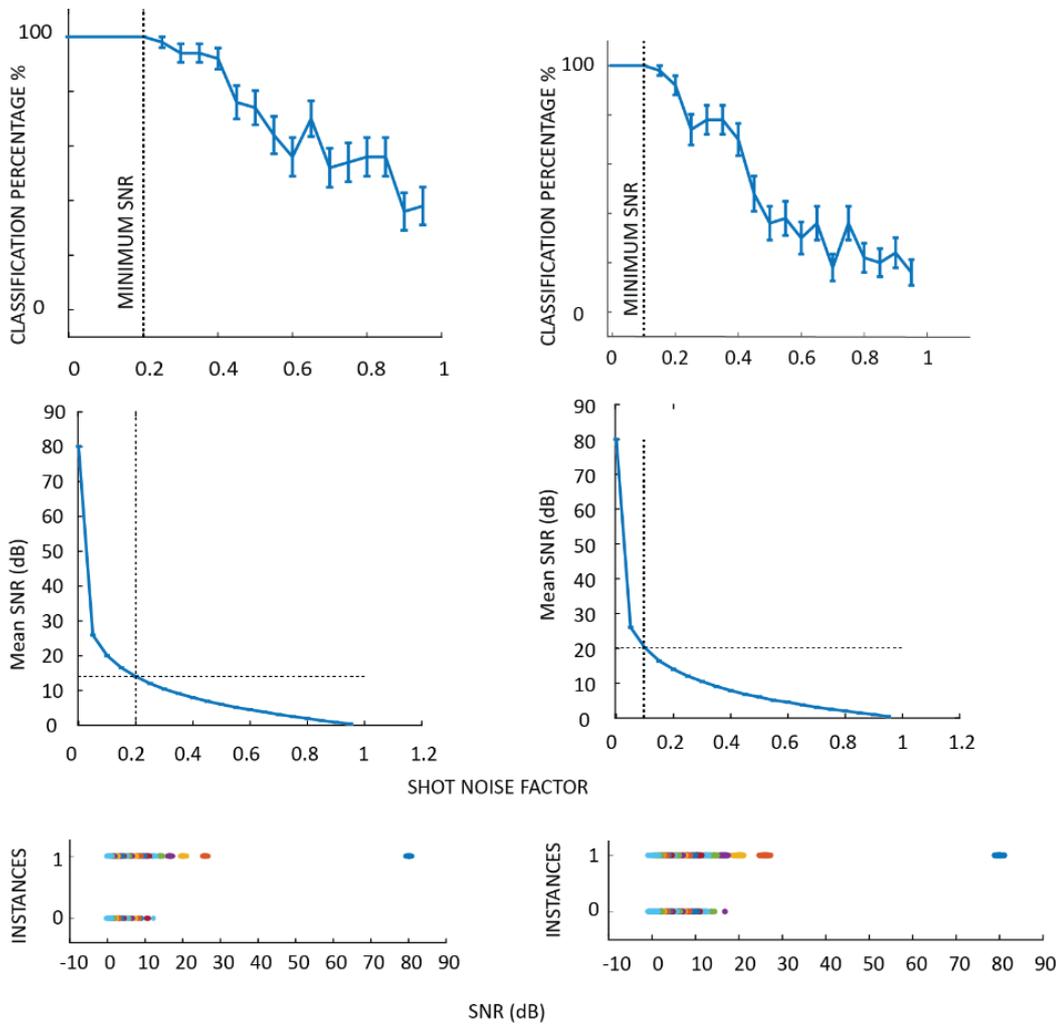

**Figure S2:** Minimum required SNR (Min. SNR) requirements for lateral (*Left panel*) and axial (*Right panel*) separation case in shot-noise DNVMM maps. **(A)** *Left* Plot of correct classification percentage versus shot noise factor (see Methods, SNR Calculations) for lateral separation case (two neurons ,10$\mu m$ apart, spike time difference 0.5$ms$). Min. SNR has been marked as the point where the standard deviation of correct classification percentage drops to zero (in 50 repetitions) **(A)** *Right* Plot of current classification percentage versus shot noise factor for axial separation case (two neurons, 7$\mu m$ apart, spike time difference 0.5$ms$). Vertical line marks point of Min. SNR. **(B)** *Left* Plot of mean SNR (dB) versus shot noise factor for lateral separation case. **(B)** *Right* Plot of mean SNR (dB) versus shot noise factor for axial separation case. (A) and (B) plots share same X axis title– Shot noise factor. **(C)** *Left* Example individual instances of correct(1)/ incorrect(0) reconstruction by proposed algorithm versus SNR(dB) for lateral separation case. A clear shift in certainty of reconstruction is observed at SNR above Min. SNR. **(C)***Right* Example individual instances of correct(1)/ incorrect(0) reconstruction by proposed algorithm versus SNR(dB) for axial separation case. Left and Right panels of (C) share same X axis title – SNR. Different colors are used to distinguish different individual instances.

Supplementary Table S1

Performance in case of dictionary elements based on time points 3.5ms, 4ms, 4.5ms (Fig. 2)

| Simulation Type | Actual spike instances (a) | Algo marked spike instances (b) | Correctly Marked by Algorithm (c=intersect (a, b)) | Percentage performance c/bX100 | Percentage performance (divided into 10 bins) Mean+/-std error | SNR (dB) | f |
|---|---|---|---|---|---|---|---|
| 2D without noise | 579 | 456 | 382 | 83.77 | 83.61 +/- 2.17 | -- | 0.994 |
| 2D with noise | 579 | 474 | 396 | 83.54 | 83.82+/- 2.09 | -11.87dB | 0.994 |
| 3D without noise | 1727 | 898 | 612 | 68.15 | 67.77+/-1.41 | -- | 0.9996 |
| 3D with noise | 1727 | 727 | 523 | 71.94 | 71.73+/-1.19 | -9.46 dB | 0.9996 |

Supplementary Table S2

Performance in case of dictionary elements based on time points 3ms, 3.5ms and 4ms (Fig. 2)

| Simulation Type | Actual spike instances (a) | Algo marked spike instances (b) | Correctly Marked by Algorithm (c=intersect (a, b)) | Percentage performance c/bX100 | Percentage performance (divided into 10 bins) Mean+/-std error | SNR (dB) | f |
|---|---|---|---|---|---|---|---|
| 2D without noise | 579 | 499 | 425 | 85.17 | 85.45+/-1.37 | -- | 0.994 |
| 2D with noise | 579 | 686 | 389 | 56.71 | 57.12+/-2.04 | -11.8375 | 0.994 |
| 3D without noise | 1727 | 63 | 33 | 52.38 | 40.39+/-9.84 | -- | 0.9996 |
| 3D with noise | 1727 | 903 | 2 | 0.22 | 0.23+/- 0.22 | -9.4641 | 0.9996 |